\documentclass[epj,nopacs]{svjour}
\usepackage{graphics}
\begin{document}
\title{Initial-state parton shower kinematics for NLO event generators}
\author{Shigeru Odaka \thanks{E-mail: shigeru.odaka@kek.jp}
and  Yoshimasa Kurihara
}                     
%
%
\institute{High Energy Accelerator Research Organization (KEK),
        1-1 Oho, Tsukuba, Ibaraki 305-0801, Japan}
\date{Received: date / Revised version: date}
%
\abstract{
We are developing a consistent method to combine tree-level event generators 
for hadron collision interactions 
with those including one additional QCD radiation 
from the initial-state partons, 
based on the limited leading-log (LLL) subtraction method, 
aiming at an application to NLO event generators. 
In this method, a boundary between non-radiative and radiative processes 
necessarily appears at the factorization scale ($\mu_{F}$).
The radiation effects are simulated using a parton shower (PS) in 
non-radiative processes.
It is therefore crucial in our method to apply a PS which well reproduces 
the radiation activities evaluated from the matrix-element (ME) calculations 
for radiative processes. 
The PS activity depends on the applied kinematics model.
In this paper 
we introduce two models for our simple initial-state leading-log PS: 
a model similar to the "old" PYTHIA-PS and a $p_{T}$-prefixed model 
motivated by ME calculations. 
PS simulations employing these models are tested using 
$W$-boson production at LHC as an example.
Both simulations show a smooth matching to the LLL-subtracted 
$W$ + 1 jet simulation in the $p_{T}$ distribution of $W$ bosons, 
and the summed $p_{T}$ spectra are stable against a variation of $\mu_{F}$,
despite that the $p_{T}$-prefixed PS results in an apparently harder 
$p_{T}$ spectrum.
%
} 
\authorrunning{S. Odaka and Y. Kurihara}
\maketitle

\onecolumn


\section{Introduction}
Next-to-leading order (NLO) corrections in Quantum Chromodynamics (QCD) 
are required by experiments 
for precise measurements of known processes 
and reliable estimates of the background for expected heavy 
particle productions at high-energy hadron collisions, 
such as those at Fermilab Tevatron and the forthcoming CERN LHC. 
The corrections are desired to be implemented in Monte-Carlo event 
generators because they may depend on the topology 
of the events experimentally selected.

The implementation of NLO corrections in event generators, in general, 
has a problem in regularizing the collinear and soft divergences; 
{\it i.e.}, the cancellation of divergences in radiative processes 
and non-radiative processes.
Phase-space slicing methods that are usually applied in electroweak 
corrections are problematic for QCD corrections because of the 
large coupling.
They frequently lead to large negative cross sections 
in non-radiative processes 
and/or visible boundaries in the phase space of radiative processes.
These properties are undesirable for use in experimental analyses.

For estimating interactions at high-energy hadron collisions, 
it is also necessary to employ an appropriate parton distribution 
function (PDF) and fragmentation functions (FF), 
depending on the energy scale of the perturbatively evaluated 
hard interaction. 
They resum large logarithmic collinear corrections to all orders
to improve the convergence of the perturbation. 
These collinear corrections are usually simulated in event generators 
in the form of a parton shower (PS), 
because experiments require an exclusive simulation of beam collisions, 
and the corrections produce an experimentally visible transverse boost 
of the hard interaction and additional hadron activities. 
The corrections by PDF, FF and PS have an apparent overlap 
with real emissions in NLO corrections 
based on matrix-element (ME) calculations.
We have to avoid double counts in order to construct consistent 
NLO event generators.

Subtraction methods \cite{catani,collins} naturally solve these problems, 
in which the divergent terms are subtracted from radiative processes 
and their integrated contributions are added to non-radiative processes.
The emergence of negative cross sections is limited 
and all results become finite, 
since the divergences are canceled within non-radiative processes.
The application of PDF/FF and/or PS automatically restores the corrections, 
if the subtracted terms are identical to their leading contributions.
The concept is simple, 
while implementation in event generators is not easy 
because the matching between the subtraction and the actual PS 
implementation is not trivial.
After many trials, a practical solution \cite{mcatnlo} has been presented 
based on detailed knowledge on the initial-state PS of HERWIG \cite{herwig}, 
and has been applied to several color-singlet production 
and heavy-flavor production processes where it is enough to consider 
radiations from initial-state partons.
Recently, a new idea has been proposed to make the cross sections 
totally positive \cite{nason}.

We are developing NLO event generators based on a different concept 
of subtraction \cite{kurihara}. 
Our goal is to develop an automated generation system of NLO event 
generators for hadron collision interactions based on GRACE \cite{grace}.
We adopt a leading-log (LL) subtraction method 
for the matching between real emissions and PDF/PS, 
where a naive LL approximation to the parton radiation, 
proportional to $1/Q^{2}$, is subtracted from radiative processes.
Though the implementation is still limited to the initial-state radiation, 
extension to the final state will be easy since the concept is quite simple.

In our method, radiative processes simulate the hard LL part, harder than 
the factorization scale ($\mu_{F}$), and the non-LL part of the radiation.
The softer LL part is simulated by the PS applied to non-radiative processes.
Therefore, it is crucial in this method to employ a PS that well matches 
with the ME calculation of the radiation.
In order to make the discussions transparent,  
we have developed an initial-state PS that strictly reproduces theoretical 
arguments in the LL approximation.
A primary feasibility of this method has been demonstrated in a previous 
paper \cite{kurihara}.
The theoretical arguments are, however, performed only at the collinear limit.
We need to introduce an appropriate kinematics model in order to construct 
a PS that is usable in practical event generators.
The model applied in the demonstration was very primitive.

In the present study we introduce two practical models of the PS kinematics 
and examine their matching properties, 
using $W$-boson production at the LHC condition as an example.
Since this process has an apparent fixed energy scale of 
the $W$-boson mass ($m_{W}$), 
possible mismatches would become visible in physical distributions.
The matching can be tested without introducing loop corrections.
The tests are carried out in the present study 
by using simulations employing tree-level ME calculations 
of the inclusive $W$ ($W$ + 0 jet) and $W$ + 1 jet production processes,
where "jet" denotes a light quark or gluon in the final state.
We examine only the internal consistency of our method.
Comparison with experimental data will be done in a separate paper, 
where additional lower $Q^{2}$ effects, 
such as the intrinsic $p_{T}$ and the underlying events, 
must be taken into account.

We need to include loop corrections, 
together with cancelling soft and collinear corrections, 
in order to construct complete NLO event generators.
Inclusion of these corrections will result in a modification 
to the normalization of non-radiative processes.
Accordingly, it will produce a certain mismatch at the boundary ($\mu_{F}$), 
even with a perfect matching method at the tree level.
We will need to apply an appropriate modification to radiative processes, 
if we want to restore a smooth matching.
By the way, this is merely a technical issue 
since this mismatch is a quantity at the level of NNLO 
(next-next-to-leading order), 
beyond the scope of NLO corrections.
In addition we have a plan to construct a full NLL (next-to-leading log) 
parton-shower program \cite{nll-ps}, 
in order to achieve a rigorous theoretical matching at the NLO level. 
The subtraction will have to be changed accordingly if it becomes available.

This paper is organized as follows: the bases of our LL-PS and the LL 
subtraction are given in Section 2. 
The description includes some details of the LL subtraction for 
the sample process.
The kinematics models are introduced and tested in Sections 3 and 4.
Finally, the conclusion is presented in Section 5.

\section{Parton Shower and LL subtraction}

We use an $x$-deterministic forward evolution \cite{kurihara} 
for the initial-state PS.
This technique is a solution to overcome the low-efficiency problem 
in the forward evolution. 
Starting from a small $Q^{2}$, 
the PS strictly reproduces the QCD evolution implemented in PDFs.
The final Bjorken's $x$ is thus given by the PS.
The PS energy scale ($\mu_{PS}$), the maximum energy scale in PS, 
is therefore necessarily identical to the factorization scale ($\mu_{F}$), 
the energy scale of PDF.
This means that the PS radiation is limited by $\mu_{F}$.
The limitation is taken into account in the subtraction as well. 
Hereafter, we call this technique the limited leading-log (LLL) subtraction.

In our PS we refer to a PDF at a low energy scale ($Q_{0}$ = 4.6 GeV) 
and produce branches in increasing order of $Q^{2}$ 
until reaching $Q^{2} = \mu_{F}^{2}$.
The $Q^{2}$ of each branch is determined according to the Sudakov 
form factor, expressed as 
\begin{equation}\label{sudakov1}
	S(Q_{1}^{2}, Q_{2}^{2}) = \exp\left[ - \int_{Q_{1}^{2}}^{Q_{2}^{2}}
	{dQ^{2} \over Q^{2}} \int_{\epsilon}^{1-\epsilon} dz \  
	{\alpha_{s}(Q^{2}) \over 2\pi}\ P(z) \right] ,
\end{equation}
where $P(z)$ is the Altarelli-Parisi splitting function at the leading order; 
\begin{eqnarray}
	P(z)_{q \rightarrow qg} & = & C_{F} {1+z^{2} \over 1-z}, \\
	P(z)_{g \rightarrow gg} & = & N_{C} {(1-z(1-z))^{2} \over z(1-z)}, \\
	P(z)_{g \rightarrow q \bar{q}} & = & T_{R} ( z^{2} + (1-z)^{2}), 
\end{eqnarray}
for the branches $q \rightarrow qg$, $g \rightarrow gg$ and 
$g \rightarrow q \bar{q}$, respectively, 
with $C_{F} = 4/3$, $N_{C} = 3$ and $T_{R} = n_{f}/2 = 5/2$.
The splitting function $P(z)$ is summed over possible branches. 
We use {\tt CTEQ5L} \cite{cteq5} for the reference PDF in the present study.
Accordingly we use the first-order strong coupling, 
\begin{equation}\label{alpha_s}
	\alpha_{s}(Q^{2}) = { 4\pi \over \beta_{0}\ln({Q^{2}/\Lambda^{2}) } }
\end{equation}
with $\beta_{0} = 11 - 2n_{f}/3$ and $\Lambda$ = 0.146 GeV for $n_{f} = 5$.
The branching mode and the parameter $z$ are determined in proportion to 
$P(z)$ of relevant branches. 
An unphysical parameter, $\epsilon$, is introduced 
in order to cutoff the divergence of $P(z)$.
We set $\epsilon = 10^{-6}$ as the default.
Physical properties do not depend on this choice 
if $\epsilon$ is set to be sufficiently small.

PS is a Monte-Carlo solution of the QCD evolution 
based on the factorization theory.
In this theory, the evolution is considered at the collinear limit 
in an infinite-momentum frame. 
The transverse behavior of the radiation is given only 
at the first-order approximation. 
As a result, the predicted kinematics do not strictly conserve 
the energy and momentum.
We therefore need to introduce a certain model of the branching kinematics 
that gives a correspondence of the above branching parameters, 
$Q^{2}$ and $z$, to kinematical variables in a finite-momentum frame, 
to construct a practical PS strictly conserving the energy and momentum.
The purpose of this paper is to introduce such models 
and test their feasibility in our subtraction method.
The use of our naive PS program allows us to clearly separate 
this model-dependent part from theoretically well-defined parts.

Though the LL subtraction is not a main subject of this paper, 
we here describe some details for the present example process, 
$W$ + 1 jet production, 
since it is used for the tests of PS models in the following sections.
At the parton level, the $W$ + 1 jet production process is composed of 
three incoherent subprocesses:
\begin{equation}\label{qqbar}
	q \bar{q}' \rightarrow g W,
\end{equation}
\begin{equation}\label{gqbar}
	g \bar{q}' \rightarrow q W,
\end{equation}
\begin{equation}\label{qg}
	q g \rightarrow q' W.
\end{equation}
We ignore CKM nondiagonal interactions in the present study.
Therefore, $(q, q')$ denotes $(d, u)$, $(u, d)$, $(s, c)$ or $(c, s)$.

The $W$ bosons are assumed to decay to a pair of an electron 
and a neutrino.
The matrix elements are evaluated including this decay at the tree level, 
assuming a total decay width of 2.12 GeV.
The calculations are performed within the framework of BASES/SPRING 
\cite{bases} employing the matrix-element codes generated 
by GRACE \cite{grace}. 
The studies in this paper are carried out using cross-section results 
from BASES.
The same method is applied for simulating the inclusive $W$ ($W$ + 0 jet) 
production, too.

Subtraction is done at the matrix-element level as \cite{kurihara}
\begin{equation}\label{LLsub}
	\vert {\cal M}_{\rm LLsub}(\hat{s}; \mu_{R}) \vert ^{2} = 
	\vert {\cal M}_{W+j}(\hat{s}; \mu_{R}) \vert ^{2} 
	- \vert {\cal M}_{W}(\hat{s}_{W}) \vert ^{2} f_{\rm LL}(x,t; \mu_{R}), 
\end{equation}
where ${\cal M}_{W+j}$ is the matrix element for a given $W$ + 1 jet event 
with a squared center-of-mass (cm) energy of $\hat{s}$, 
and ${\cal M}_{W}$ is that for the $W$ production sub-system in this event 
having a squared invariant mass of $\hat{s}_{W}$.
The rotation of the sub-system is also taken into account in the calculation, 
though it is irrelevant to the following studies 
since the angular properties are always integrated.
The LL radiation factor $f_{\rm LL}$ can be written as
\begin{equation}\label{LLfactor}
	f_{\rm LL}(x,t; \mu_{R}) = {\alpha_{s}(\mu^{2}_{R}) \over 2\pi}\ 
	{P(x) \over x} {16 \pi^{2} \over |t|}, 
\end{equation}
where $P(x)$ is an Altarelli-Parisi splitting function relevant 
to this event.
It is essential to use an identical definition of $\alpha_{s}$ 
in the $W$ + 1 jet matrix element ${\cal M}_{W+j}$ 
and in the LL factor $f_{\rm LL}$. 
We use the definition of Eq. (\ref{alpha_s}) with the energy scale $Q^{2}$ 
set to a fixed value ($\mu^{2}_{R}$) in the present study.

The parameter $x$ is the ratio of the squared cm energies 
($x = \hat{s}_{W}/\hat{s}$), 
and $t$ is the squared momentum transfer of the parton branch 
in the picture of PS.
There are two possibilities in the definition of $t$ 
for the subprocess (\ref{qqbar}).
The gluon in the final state can be radiated from $q$ and $\bar{q}'$ 
in the initial state.
We have to subtract both contributions.
Thus, $t$ is defined using the four-momenta of the partons 
as $1/|t| = 1/|t_{q}| + 1/|t_{q'}|$ 
with $t_{q} = (p_{q} - p_{g})^{2}$ and $t_{q'} = (p_{q'} - p_{g})^{2}$.
On the other hand, there is no ambiguity for subprocesses (\ref{gqbar}) 
and (\ref{qg}), 
since $g \rightarrow q \bar{q}$ is the unique possible branch.
The parameter $t$ is defined as $t = (p_{g} - p_{q})^{2}$ for (\ref{gqbar}) 
and $t = (p_{g} - p_{q'})^{2}$ for (\ref{qg}).

\begin{figure}
\begin{center}
\resizebox{0.8\textwidth}{!}{
  \includegraphics{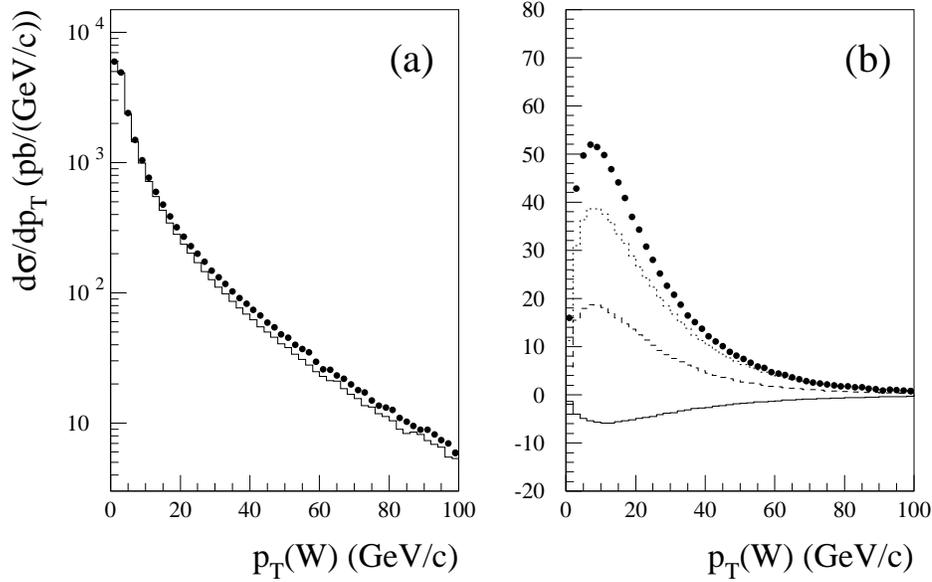}
}
\caption{
\label{fig_llsub}
$W$ + 1 jet cross section at LHC as a function of the $W$-boson $p_{T}$:
(a) the sum of three incoherent subprocesses, 
and (b) the residual of the LL subtraction.
A $p_{T}$ cut of 1 GeV/$c$ is applied, while PS is yet to be applied 
in these results.
The result from exact matrix elements at the tree level (filled circles) 
is compared with that from the LL approximation (histogram) in (a).
The result is shown for each subprocess, $q \bar{q}' \rightarrow g W$ 
(solid), $g \bar{q}' \rightarrow q W$ (dashed), 
and $q g \rightarrow q' W$ (dotted), separately in (b), 
together with the sum shown with filled circles.
}
\end{center}
\end{figure}

The subtraction result at the cross-section level is shown 
in Fig. \ref{fig_llsub} 
as a function of the transverse momentum of the $W$ boson, 
where PS is yet to be applied.
The matrix elements are converted to the cross sections assuming 
the LHC condition (proton-proton collisions at the cm energy of 14 TeV) 
and using {\tt CTEQ5L} for PDF.
The renormalization scale $\mu_{R}$ and the factorization scale $\mu_{F}$ 
are both fixed to the $W$-boson mass ($m_{W}$ = 80.2 GeV).
A $p_{T}$ cut is necessary to apply 
since the subtraction is done numerically.
We set the minimum $p_{T}$ to 1 GeV/$c$.
These conditions are valid throughout the present study, 
except for the choice of $\mu_{F}$.

Figure \ref{fig_llsub}(a) shows the cross section based on the exact 
$W$ + 1 jet matrix element (filled circles) and that based on the LL 
approximation (histogram); 
{\it i.e.}, they correspond to the first and the second term 
in the right-hand side of Eq. (\ref{LLsub}), respectively.
The result from the subtracted matrix element is plotted 
in Fig. \ref{fig_llsub}(b) with filled circles.
The result is also shown for each subprocess separately.
The residual is always negative for subprocess (\ref{qqbar}) 
and positive for (\ref{gqbar}) and (\ref{qg}).
The sum (filled circles) is positive since the contribution 
of the subprocess (\ref{qg}) is large in proton-proton collisions.
We can see that the residual is going to vanish as $p_{T}$ goes to zero 
in all subprocesses, 
and the small $p_{T}$ cut does not significantly affect the final results.

In the LLL (limited LL) subtraction, 
the subtraction is applied only when $|t_{q}| < \mu^{2}_{F}$ or 
$|t_{q'}| < \mu^{2}_{F}$ in the subprocess (\ref{qqbar}), 
and when $|t| < \mu^{2}_{F}$ in the subprocesses (\ref{gqbar}) 
and (\ref{qg}). 
Here we note that both the $W$ + 0 jet and the subtracted $W$ + 1 jet 
cross sections depend on $\mu_{F}$.
The LL contributions of the parton radiation below the energy scale 
of $\mu_{F}$ are resummed by PDF and PS in the former, 
while those above $\mu_{F}$ are explicitly evaluated 
at the first order of $\alpha_{s}$ in the latter.
The non-LL contributions are all remaining in the latter.
Thus, we can expect that the dependence on $\mu_{F}$ cancels 
in the sum of these two cross sections, 
at least at the leading order of $\alpha_{s}$.

\begin{figure}
\begin{center}
\resizebox{0.6\textwidth}{!}{
  \includegraphics{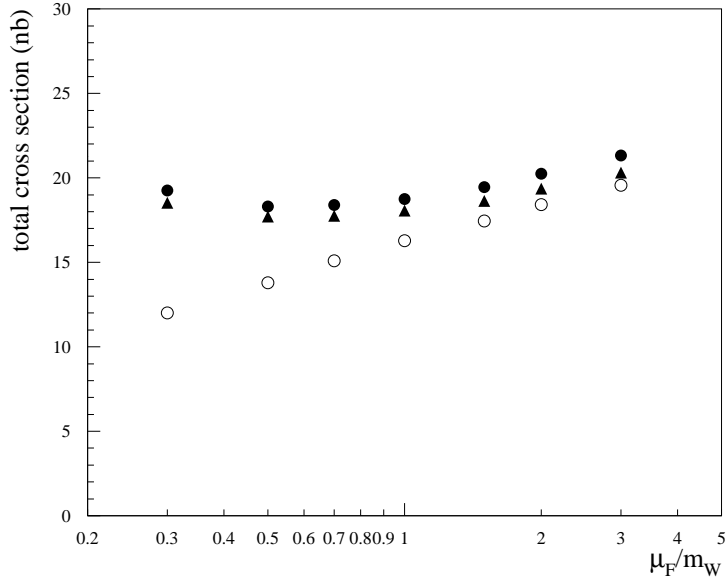}
}
\caption{
\label{fig_scale_dep}
Factorization scale ($\mu_{F}$) dependence of the total cross section 
for $W$ production at LHC.
An apparent $\mu_{F}$ dependence in the inclusive $W$ production 
cross section (open circles) is greatly reduced 
if the $W$ + 1 jet cross section with the LLL (limited LL) subtraction 
is added (filled circles). 
Filled triangles illustrate the results from a simulation where the PS 
employing the $p_{T}$-prefixed kinematics model is applied.
The cross section result is affected by the PS veto in the $W$ + 1 jet 
simulation.
The PDF is slightly changed since the QCD evolution is evaluated by PS 
in our simulation.
}
\end{center}
\end{figure}

The summed total cross section is plotted in Fig. \ref{fig_scale_dep} 
with filled circles as a function of $\mu_{F}$.
They should be compared with the inclusive $W$ production cross section 
shown with open circles.
We can see that the $\mu_{F}$ dependence is greatly reduced 
in the summed cross section, as we expected.
The variation is at most 10\% within the range of 
$1/2 \leq \mu_{F}/m_{W} \leq 2$.
This is a remarkable feature of the LLL subtraction.
The stability will become better if an appropriate resummation 
(Sudakov suppression) is applied to the $W$ + 1 jet process.
It will suppress a small enhancement at very small $\mu_{F}$ values 
in Fig. \ref{fig_scale_dep}, 
as demonstrated in a previous paper \cite{odaka}.

\section{PYTHIA kinematics}

We first test the kinematics model of the "old" PYTHIA-PS 
for the initial state \cite{pythia-isr}
because this PS is constructed on nearly the same theoretical basis as ours.
The starting assumption of this model is that the $Q^{2}$ of the evolution 
is identical to the virtuality of the evolving parton, 
\begin{equation}\label{assumption1}
  p_{\rm evol}^{2} = -Q^{2} ,
\end{equation}
and $z$ gives the ratio between the squared cm energies after and before 
a branch, 
\begin{equation}\label{assumption2}
  z = \hat{s}^{\prime}/\hat{s} .
\end{equation}
The latter ensures the relation $\hat{s}_{\rm hard} = x_{1}x_{2}s$ 
between the squared cm energies of the hard interaction and 
the beam collision, 
with $x$ given by the product of all $z$ values in each beam.

\begin{figure}
\begin{center}
\resizebox{0.5\textwidth}{!}{
  \includegraphics{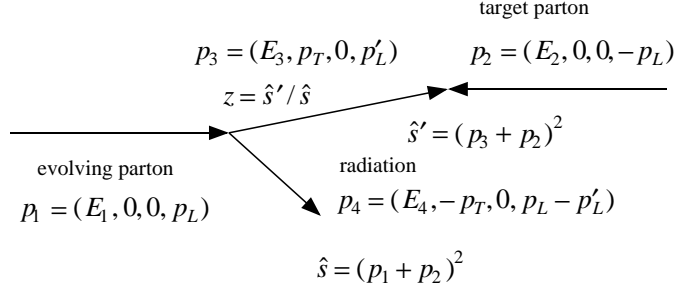}
}
\caption{
\label{fig_kinem}
Definition of kinematical variables in each parton branch.
The calculation is done in the cm frame of the evolving parton 
and the target parton.
}
\end{center}
\end{figure}

The transverse momentum ($p_{T}$) of each branch is calculated 
from energy-momentum conservation conditions \cite{pythia-isr,odaka}.
The definition of kinematical variables is illustrated 
in Fig. \ref{fig_kinem}.
Assumption (\ref{assumption1}) leads to $p^2_{3} = -Q^{2}$, 
while $p^2_{1}$ is given by the previous branch.
The definition of $z$ requires a target parton.
The definition is not trivial since two incoming partons are both evolving.
We execute the branches in the increasing order of $Q^{2}$ 
equally taking both sides into consideration.
The remaining parton on the opposite side is chosen as the target.
Therefore, the target parton is also virtual.
Going into further details, the calculation sometimes gives 
a negative $p_{T}^{2}$.
In such cases we set $p_{T}$ to zero and adjust the $Q^{2}$ to give 
this solution. 
These calculations are done in the cm frame of the evolving parton and 
the target parton as shown in Fig. \ref{fig_kinem}.
The momenta are rotated and boosted back to the cm frame 
of the starting partons 
every time after the branching kinematics are determined.
After completing all branches, the system is boosted to the laboratory 
(beam-collision) frame, 
and the hard interaction part is attached according to the momenta of 
finally remaining two partons.

\begin{figure}
\begin{center}
\resizebox{0.8\textwidth}{!}{
  \includegraphics{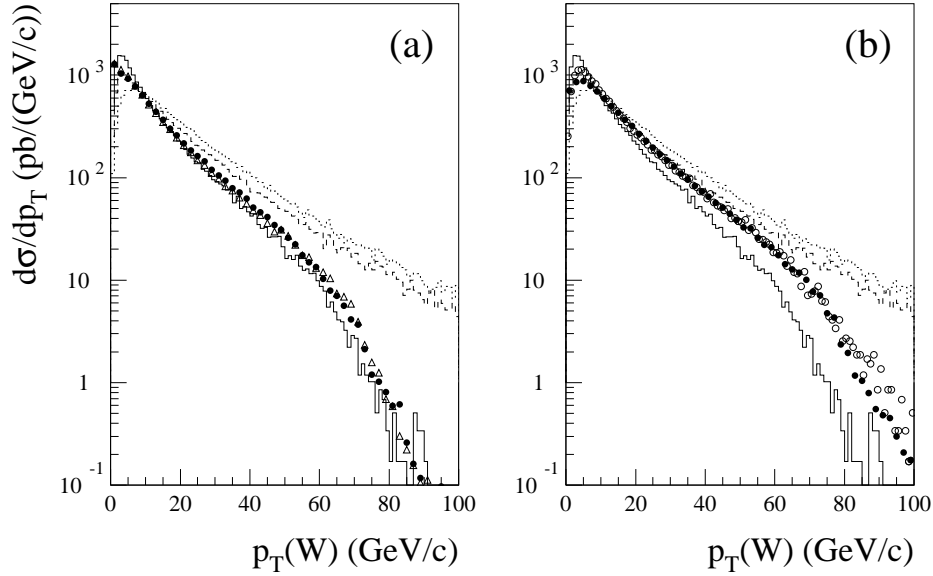}
}
\caption{
\label{fig_w0j}
Simulated transverse momentum ($p_{T}$) distribution of $W$ bosons 
in the inclusive $W$ production process at the LHC condition.
Filled circles show the result of our simulation: 
(a) with the "old" PYTHIA-PS kinematics model 
and (b) with the $p_{T}$-prefixed model.
Solid histograms show the result from "old" PYTHIA without the ME 
correction to the initial-state radiation.
The "old" and "new" PYTHIA simulations with the ME correction are 
illustrated with dashed and dotted histograms, respectively, for reference.
Open triangles in (a) show the result of a $p_{T}$-prefix model 
with a non-standard assumption of $p^{2}_{T} = (1-z)^{2}Q^{2}$, 
and open circles in (b) the result of the "new" PYTHIA simulation 
without the ME correction.
}
\end{center}
\end{figure}

Filled circles in Fig. \ref{fig_w0j}(a) show the simulated $p_{T}$ 
distribution of $W$-bosons in the inclusive $W$ production process 
for the LHC condition with $\mu_{F} = m_{W}$.
The corresponding prediction from the "old" PYTHIA-PS is shown 
with a histogram, 
where PYTHIA 6.403 \cite{pythia64} is used with {\tt MSTP(68) = 0}, 
{\tt MSTP(71) = 0}, {\tt MSTP(81) = 0} and {\tt MSTP(111) = 0}; 
{\it i.e.}, the ME-correction, final-state PS, multiple-interaction 
and hadronization are turned off. 
The PDF choice is the default; namely, {\tt CTEQ5L} is used, too.
We can see a good agreement between these two simulations.
Any difference in a very low $p_{T}$ region ($<$ 10 GeV/$c$) should be 
ignored because our $Q_{0}$ is relatively large and no intrinsic $p_{T}$ 
is considered.
In the figure, PYTHIA predictions with the ME correction 
({\tt MSTP(68) = 3}) \cite{pythia-mecor} 
using the "old" PS ({\tt MSTP(81) = 0}) and the "new" PS 
({\tt MSTP(81) = 20}) are also illustrated for a reference.

\begin{figure}
\begin{center}
\resizebox{0.8\textwidth}{!}{
  \includegraphics{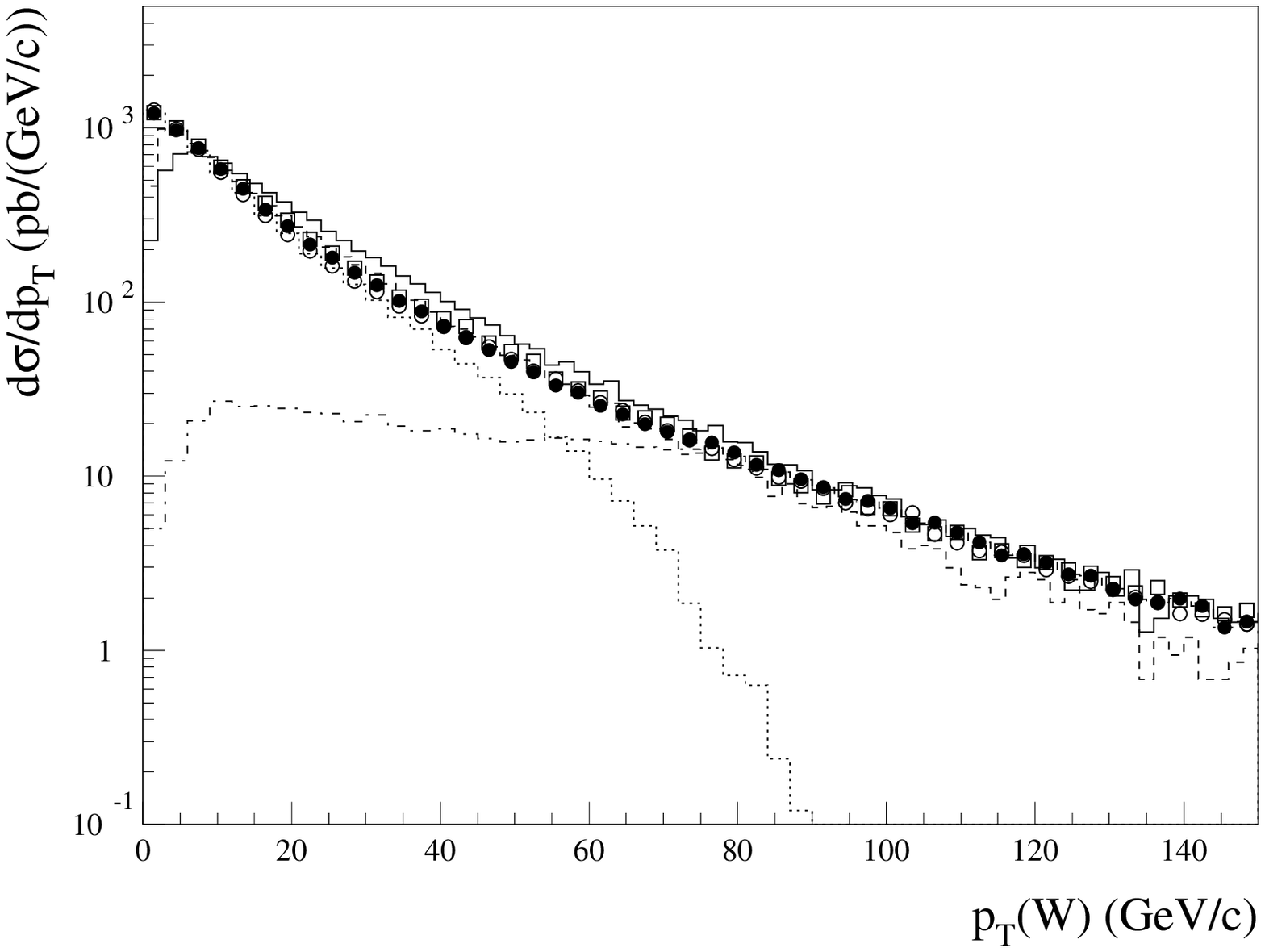}
}
\caption{
\label{fig_py-kinem}
Tree-level summation of the $W$ + 0 jet simulation and the LLL-subtracted 
$W$ + 1 jet simulation for the LHC condition 
using the PS with the "old" PYTHIA-PS kinematics model. 
The $W$ + 0 jet result for the energy-scale choice of $\mu_{F} = m_{W}$ is 
shown with a dotted histogram, the $W$ + 1 jet with a dot-dashed histogram, 
and the sum is plotted with filled circles.
The summed results for $\mu_{F} = 0.5 m_{W}$ and $\mu_{F} = 1.5 m_{W}$ are 
plotted with open circles and open squares, respectively, 
in order to show the stability against the $\mu_{F}$ variation.
The dashed and solid histograms show the ME-corrected PYTHIA predictions 
for the $W$ production using the "old" and "new" PS models, respectively.
}
\end{center}
\end{figure}

The inclusive $W$ simulation result is also shown with a dotted histogram 
in Fig. \ref{fig_py-kinem} 
together with the result of the LLL subtracted $W$ + 1 jet simulation 
(dot-dashed histogram) described in the previous section, 
with the same energy-scale choice of $\mu_{F} = m_{W}$.
The PS is also applied to the $W$ + 1 jet simulation with a PS veto, 
where the PS is retried if the maximum $p_{T}$ of the PS branches exceeds 
the $p_{T}$ of the $W$ + 1 jet ME.
The sum of these results is plotted with filled circles.
We can see a smooth transition between the two simulations.
As described in the previous section, 
the summed cross section shows good stability against the variation 
of $\mu_{F}$.
A similar stability should also be observed in this $p_{T}$ spectrum 
if the applied PS kinematics shows a good matching to the ME calculation.

The open circles and open squares in Fig. \ref{fig_py-kinem} show 
the summed results for other energy-scale choices of $\mu_{F} = 0.5 m_{W}$ 
and $\mu_{F} = 1.5 m_{W}$, respectively.
We can see good stability.
Though it is not shown in the figure, 
the result for $\mu_{F} = 2.0 m_{W}$ shows a visible discrepancy 
from these results. 
This is reasonable since the PS is constructed on the basis of a theory 
and a model relevant to soft radiations.
The underlying approximations may become inappropriate for hard radiations 
where the hardness ({\it e.g.}, $p_{T}$) is comparable to or larger than 
the typical energy scale of the considered hard interaction, $m_{W}$ 
in the present case.
The contribution of such hard radiations may become significant 
if we set $\mu_{F}$ to be larger than $m_{W}$.
The energy scale $\mu_{F}$ should not be set very large in our method.

Although the resultant simulation is self-consistent, 
the summed $p_{T}$ spectrum seems to be rather soft in low to medium 
$p_{T}$ regions ($<$ 50 GeV/$c$) compared to the ME-corrected simulations 
of PYTHIA.
This is because the applied PS is soft, as we can see 
in Fig. \ref{fig_w0j}(a).
Though this may be reasonable, 
we try to construct another kinematics model giving a harder spectrum 
in the next section.

It may be worth noting here that the two ME-corrected PYTHIA simulations 
disagree with each other, 
as we can see in Figs. \ref{fig_w0j} and \ref{fig_py-kinem}.
The simulation with the "new" PS is harder and in good agreement 
with our simulation in the high-$p_{T}$ region.
Since the high-$p_{T}$ spectrum is predominantly determined 
by the unambiguous $W$ + 1 jet ME in our simulation, 
the "new" PS simulation must be reasonable.
The softness of the "old" PS simulation that we see here may also be 
related to its basic property, 
which is discussed in the next section.

\section{$p_{T}$-prefixed kinematics}

As we have shown in a previous report \cite{odaka}, 
the kinematics of the "old" PYTHIA-PS leads to the relation 
\begin{equation}\label{pt1}
  p_{T}^{2} = (1-z)^{2}Q^2 
\end{equation}
in each branch at the soft limit. 
As we can see in Eq. (\ref{sudakov1}), the Sudakov form factor 
becomes smaller 
if we choose a smaller value for the unphysical parameter, $\epsilon$.
This means that the number of parton branches increases. 
Thus, if we make $\epsilon$ sufficiently small, 
the $Q^{2}$ step of the PS evolution becomes nearly continuous compared 
to the energy scale of the hard interaction.
However, the increased branches are predominantly very soft ($z \sim 1$) 
and do not affect physical properties, 
such as the $p_{T}$ of $W$ bosons and visible jets.
The above relation can be obtained at this continuous limit; 
{\it i.e.}, $p^{2}_{1} \rightarrow p^{2}_{3} = -Q^{2}$ 
using the notation in Fig. (\ref{fig_kinem}).

On the other hand, ordinary discussions concerning the branching kinematics 
give us a different relation, 
\begin{equation}\label{pt2}
  p_{T}^{2} = (1-z)Q^2 .
\end{equation}
This relation can be obtained by assuming that the incoming partons 
and the radiation are all massless; 
{\it i.e.} $p^{2}_{1} = p^{2}_{2} = p^{2}_{4} = 0$.
Apparently Eq. (\ref{pt1}) gives smaller $p_{T}$ values 
than Eq. (\ref{pt2}) since $z < 1$.
In ME calculations, all initial- and final-state partons are necessarily 
on-shell; namely, they are nearly massless.
Therefore, if we want to achieve a good matching between the PS radiation 
and the ME radiation, 
the relation (\ref{pt2}) must be better than (\ref{pt1}).
We test such a model leading to the relation (\ref{pt2}) in this section.

Although there may be some sophisticated models, 
here we adopt a straightforward way to realize this relation.
We prefix the $p_{T}$ of each branch according to Eq. (\ref{pt2}) 
as a starting assumption for calculating the branching kinematics.
We keep the definition of $z$, Eq. (\ref{assumption2}), 
since otherwise we would need to apply undesirable energy/momentum 
corrections in the connection to ME.
As a result, we have to abandon relation (\ref{assumption1}).
The $p^{2}$ of the evolving partons is derived from energy-momentum 
conservation conditions.
This must not be troublesome, since the virtuality is not a direct 
observable and the $Q^{2}$ of the evolution is not a physically 
well-defined quantity, as we have already discussed.
In the actual implementation, 
it sometimes happens that the prefixed $p_{T}$ exceeds the kinematically 
allowed maximum. 
We decrease the $p_{T}$ value to the allowed maximum in such cases.

The filled circles in Fig. \ref{fig_w0j}(b) show the result of a simulation 
with this new PS kinematics.
The evolution to give $Q^{2}$ and $z$ is the same as in the previous 
simulation.
Apparently it gives a harder $p_{T}$ spectrum of $W$ bosons.
It should be noted here that relations (\ref{pt1}) and (\ref{pt2}) are 
obtained at the soft limit ($Q^{2} \ll \hat{s}$).
It is not trivial that the difference between these two relations is 
directly reflected in the visible $p_{T}$ spectrum of $W$ bosons.
For a confirmation, we have tried another simulation 
with the $p_{T}$-prefixed PS kinematics model, 
in which Eq. (\ref{pt1}) is assumed instead of Eq. (\ref{pt2}).
The result is shown by the open triangles in Fig. \ref{fig_w0j}(a). 
It is in good agreement with the previous result, 
confirming that the "old" PYTHIA-PS kinematics actually results 
in relation (\ref{pt1}) 
and the difference in the PS property at the soft limit is reflected in 
the visible $p_{T}$ spectrum of $W$ bosons.

\begin{figure}
\begin{center}
\resizebox{0.8\textwidth}{!}{
  \includegraphics{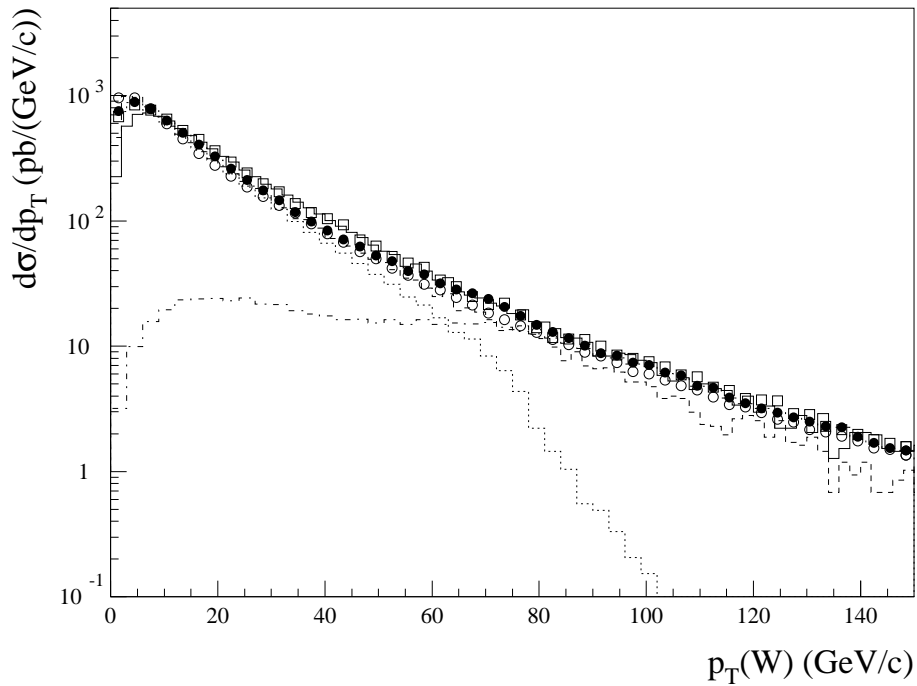}
}
\caption{
\label{fig_pt-prefix}
Tree-level summation of the $W$ + 0 jet simulation and the LLL-subtracted 
$W$ + 1 jet simulation for the LHC condition 
using the $p_{T}$-prefixed PS kinematics model. 
The $W$ + 0 jet result for the energy-scale choice of $\mu_{F} = m_{W}$ 
is shown with a dotted histogram, the $W$ + 1 jet with a dot-dashed 
histogram, and the sum with filled circles.
The summed results for $\mu_{F} = 0.5 m_{W}$ and $\mu_{F} = 1.5 m_{W}$ are 
plotted with open circles and open squares, respectively.
The dashed and solid histograms show the ME-corrected PYTHIA predictions 
for the $W$ production using the "old" and "new" PS models, respectively.
}
\end{center}
\end{figure}

The matching test is retried using the new $p_{T}$-prefixed PS 
in Fig. \ref{fig_pt-prefix}.
The simulation conditions are the same as those giving the results 
in Fig. \ref{fig_py-kinem}, except for the PS kinematics.
We can see a smooth transition between the $W$ + 0 jet and $W$ + 1 jet 
simulations again.
The spectrum in low to medium $p_{T}$ regions has become harder 
and come closer to the PYTHIA simulations with the ME correction.
The $\mu_{F}$ dependence is small, but has become larger as we can see 
in the figure.
This must be simply due to the fact that this new PS generates 
harder radiations.
Since the relations (\ref{pt1}) and (\ref{pt2}) are obtained 
at the limit where $Q^{2}/\hat{s}$ can be ignored, 
the approximation may become worse for hard radiations, 
as we have discussed in the previous section.
In fact we can see a small bump structure in the $\mu_{F}/m_{W} = 1$ 
result just below the boundary, though 
such a structure is not clear in the $\mu_{F}/m_{W} = 0.5$ result.
It is, however, possible to evaluate the internal ambiguity 
by comparing the predictions with different $\mu_{F}$ values; 
for instance, those with $\mu_{F} = m_{W}$ and $\mu_{F} = 0.5 m_{W}$. 
In general, a Sudakov suppression has to be applied 
to the $W$ + 1 jet simulation
in order to take higher-order effects into account, 
if we choose an energy scale apart from the typical energy scale 
of the considered hard interaction.
The suppression is however a few percent and can be ignored 
for $\mu_{F} = 0.5 m_{W}$.

It should be noted here again that the Bjorken's $x$ for the hard 
interaction is given by PS in our method.
The PS veto, therefore, alters the PDF for the $W$ + 1 jet simulation.
The cross section prediction slightly changes as a result.
The filled triangles in Fig. \ref{fig_scale_dep} show the results 
of a simulation with the $p_{T}$-prefixed PS 
assuming the relation (\ref{pt2}).
At present we have no idea about the question whether this alternation is 
reasonable or not.

In the course of the present study, 
we have found that the "new" $p_{T}$-ordered PS of PYTHIA shows a behavior 
quite similar to our $p_{T}$-prefixed PS. 
The result of the "new" PYTHIA-PS simulation without the ME correction is 
over-plotted in Fig. \ref{fig_w0j}(b) with open circles. 
We can see that the $p_{T}$ spectrum is almost identical to our new result.

\section{Conclusion}

In the limited leading-log (LLL) subtraction method for NLO event generators, 
the parton shower (PS) to be applied to non-radiative processes must 
well reproduce the leading-log (LL) contributions in the matrix-element (ME) 
evaluation of radiative processes.
The PS activity depends on the applied model for branching kinematics.
We have introduced two kinematics models 
to be applied to our simple initial-state leading-log PS: 
a model similar to the "old" PYTHIA-PS and a $p_{T}$-prefixed model 
with a $p_{T}$ definition motivated by ME calculations. 
The former shows a behavior similar to the "old" PYTHIA-PS, 
and the latter results in harder transverse activities, as we expected.
We also found that the "new" PYTHIA-PS is similar to the latter.

PS simulations employing these two models have been tested 
using the $W$-boson production at LHC as an example.
Both simulations show a smooth matching to the LLL subtracted $W$ + 1 jet 
simulation in the $p_{T}$ distribution of $W$ bosons, 
though the $p_{T}$-prefixed PS results in an apparently harder $p_{T}$ 
spectrum than the PS with the "old" PYTHIA kinematics.
The summed $p_{T}$ spectra are stable against the variation 
of the factorization-scale choice in both simulations. 
This is a remarkable feature of the LLL subtraction method.
The remaining instability can be used as a measure of the internal 
ambiguity of this method.

The present study is not enough to decide which model is better 
for our application.
The decision must be postponed until further studies can be conducted 
employing a comparison with experimental data, 
such as the $Z$-boson production data at Tevatron. 
Lower $Q^{2}$ effects that are absent from the present simulations has to 
be taken into account there.

The method tested in this paper is not dedicated to $W$-boson production, 
but can be applied to other processes.
Of course, $m_{W}$ has to be replaced with their {\it typical} energy scales 
in such applications.
At present, our method is limited to applications to the initial-state 
radiation.
It will, however, be easy to extend it to the final state 
since the concept is very simple.
Once extended, it will allow us to construct NLO event generators 
for those processes including jet(s) in the final state, 
which the existing NLO event generator \cite{mcatnlo} has not yet supported.
Even if loop corrections necessary for NLO calculations are absent, 
our subtraction method will give us a consistent way to simulate 
multi-jet production processes at the tree level.
This will be an alternative to the CKKW method \cite{ckkw} 
if it can be applied recursively.

\section*{Acknowledgments}

This work has been carried out as an activity of the NLO Working Group, 
a collaboration between the Japanese ATLAS group and the numerical analysis 
group (Minami-Tateya group) at KEK.
The authors wish to acknowledge useful discussions with the members: 
J. Kodaira, J. Fujimoto, T. Kaneko and T. Ishikawa of KEK, 
and K. Kato of Kogakuin University.


\end{document}